# OSDG 2.0: a multilingual tool for classifying text data by UN Sustainable Development Goals (SDGs)

**Authors**

Lukas Pukelis*, Nuria Bautista-Puig†, Gustė Statulevičiūtė*, Vilius Stančiauskas*,
Gokhan Dikmener✴, Dina Akylbekova✴

\* - PPMI; † - UC3M, ✴ - UNDP IICPSD SDG AI Lab

## Abstract

Despite concrete indicators and targets, monitoring the progress of the UN Sustainable Development Goals (SDGs) remains a challenge, given the many different actors, initiatives, and institutions involved. OSDG, an open-source classification tool aims to help navigate the SDG-related ambiguities through a simple and easy-to-use application. The tool allows to map and connect activities to the SDGs by identifying SDG-relevant content in any text. This paper presents OSDG 2.0, a new iteration of the partnership's work, which marks a significant improvement in the tool's methodology, as well as support for content in 15 languages.

## Introduction

Since their introduction in 2015, the Sustainable Development Goals (SDGs) have been guiding global sustainability efforts. Despite concrete indicators and targets, it is still a challenge to monitor the progress of the SDGs, given that many different actors, initiatives, and institutions involved. As pointed out in several studies, different interpretations of the goals may result in inaccurate narratives, which would be highly undesirable (see Armitage et al., 2020; Purnell, 2022). In 2020, the Public Policy and Management Institute (PPMI) and the United Nations Development Programme (UNDP) Istanbul International Centre for Private Sector in Development (IICPSD) SDG AI Lab launched OSDG, a tool to help the different actors: governments, companies, development organisations, universities, research funders, individual researchers, and others, to connect their activities to the SDGs (Pukelis et al., 2020).

The OSDG tool aims to help navigate SDGs related ambiguities by providing a simple and easy-to-use application that identifies SDG-relevant content in any text. In doing so, OSDG allows to map their activities or outputs to SDGs. The goals of the OSDG initiative are twofold: First, OSDG was established to tackle the duplication of efforts among many initiatives to classify content by the SDGs. The key component differentiating OSDG is the ontology, or keywords for SDG mapping. This mapping is assembled by combining various existing initiatives and approaches into a single coherent system. The project is therefore able to build upon existing research and prior efforts, as opposed to starting yet another project completely from scratch. Second, aside from integrating existing research into a comprehensive approach, the OSDG project aims to carry out the work in an open, transparent, and user-friendly manner. The acronym OSDG stands for "Open SDGs", which guides the partnership's approach. In addition to developing a transparent methodology for SDG classification, the project grants public access to the key components of the project and most of the programmatic code. OSDG is therefore accessible to anyone interested and willing to connect their work to the SDGs, especially those without access to the services of consultants and researchers.

In this paper, we present OSDG 2.0, a new iteration of the partnership's work, which marks a significant improvement in the OSDG framework. The paper is structured as follows: the first part introduces the partnership behind the project, moving on to the introduction of OSDG 2.0. The paper continues with an overview of the project's key milestones, such as holding an extensive citizen science exercise and publishing project outputs. Finally, the last part showcases existing use cases for OSDG 2.0, taken up by various project partners and stakeholders.

# 1. OSDG Partnership

OSDG is a partnership between UNDP IICPSD's SDG AI Lab and European research and policy analysis centre PPMI. The partners have joined forces to advance the research and application of frontier technologies for the achievement of the SDGs. The partnership strives to make new technologies – specifically Natural Language Processing (NLP) and text classification technologies – available to a wide range of stakeholders by advancing an open-source and free OSDG tool.

SDG AI Lab has a mission to harness the potential of novel technologies such as Artificial Intelligence (AI), Machine Learning (ML), and Geographic Information Systems (GIS) for sustainable development. The Lab provides research, development, and advisory services on the application of digital technologies to development challenges. In addition, the Lab supports UNDP's internal capacity-strengthening efforts for the increasing demand for digital solutions. To bridge the talent gap in the use of frontier technologies in development contexts, the Lab mobilises a community of volunteer data scientists, connecting UNDP teams and highly skilled data scientists to address development challenges with digital solutions.

PPMI is one of the leading public policy research centres in Europe. The institute's research and advisory role has contributed to the improvement of public policies in Lithuania and the European Union. PPMI assists its clients in dealing with contemporary public policy challenges by strengthening their administrative capacities and ability to make evidence-based strategic decisions. In addition to public policy expertise, PPMI is also highly skilled in developing various AI-based analytical applications and utilising machine learning for public policy research (Pukelis & Stančiauskas, 2019). The initial idea and implementation of OSDG stem from PPMI's work in identifying SDG-relevant context among publicly funded research projects.

# 2. OSDG 2.0

In May 2022, the OSDG team launched **OSDG 2.0**, a renewed classification tool to address the shortcomings of the previous version, and to provide an even more user-friendly experience to the general user. OSDG 2.0 has new features such as support for multiple languages, document classification and includes significant updates to the user-interface design and the backend of the OSDG platform.

The key change in OSDG 2.0 relates to a revamped methodology, namely through the incorporation of Machine Learning (ML) models. The use of ML allows to address the shortcomings of the previous version and mitigate the effects of academic and other western biases.

When the OSDG team first sought to develop a set of ML models to be used in OSDG 2.0, there was a lack of high-quality labelled data to develop models for SDG classification. While this posed significant challenges, it also provided an opportunity. It allowed the OSDG team to identify one more area to contribute to SDG research; namely by curating and opening up quality datasets, thus enabling other researchers to develop their own ML solutions. With this in mind, the team set out to launch the interactive data collection exercise, entitled the OSDG Community Platform (OSDG CP), explained in detail in the following sections.

OSDG 2.0 also goes beyond English, as many national SDG documents as well as research projects and other documents tend to be in local languages. Despite many SDG-related tools and approaches developed, English is the main vehicular language, a gap OSDG aims to tackle to respond to this emerging need. Alongside English, OSDG users can now submit content in Arabic, Danish, Dutch, Finnish, French, German, Italian, Korean, Polish, Portuguese, Russian, Spanish, Swedish, and Turkish. The team is currently working on supporting more languages, starting with Japanese and Chinese.

One of the updates for the users in the tool was the newly added support for Portable Document Format (PDF) files, while the former compatibility with the Digital Object Identifier (DOI) was removed. The ability to submit text abstracts (raw text input) remains unchanged. The upcoming versions of the OSDG tool will support other text processing formats, including rich text content.

## 2.1. Methodology

OSDG 2.0 works in two stages, presented in the scheme below.

In the current OSDG workflow, the first stage uses the ML models, trained on the data collected via the OSDG Community Platform (CP). These models carry out the initial screening of texts and suggest the preliminary SDG labels. The models also help to control the context and help to select the texts that are in the general thematic areas of the SDGs. In the second stage, OSDG uses its ontology/keyword map to verify the initial labels. The keyword matching helps to locate the 'smoking-gun' evidence that a given text is related to that particular SDG. To assign a specific SDG label, both the ML and ontology approaches must be in agreement.

**Figure 1.** Schematic workflow of OSDG 2.0

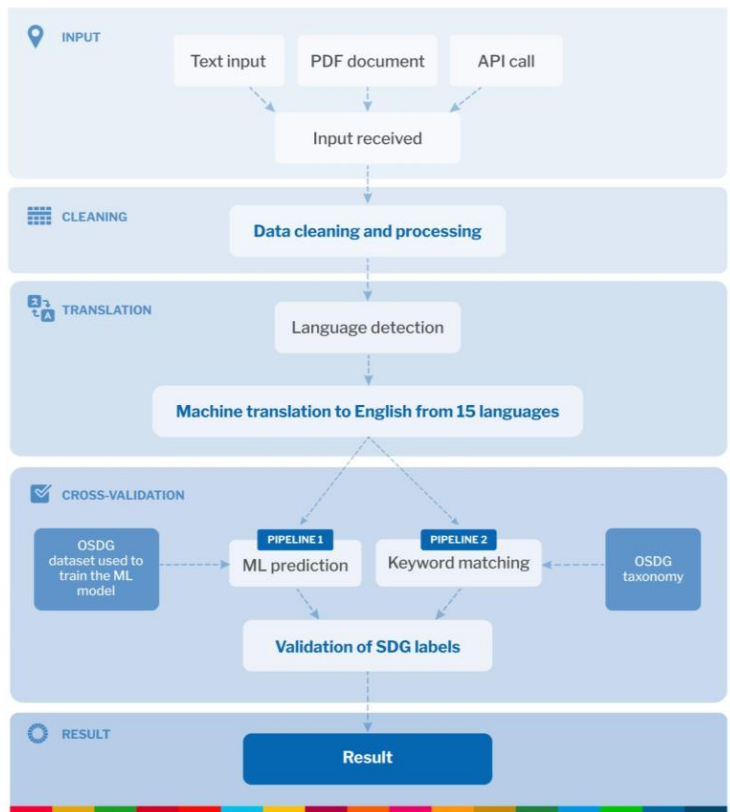

*Source: compiled by OSDG project team.*

OSDG 2.0 also introduces a data aggregation function to assign the SDG labels to longer documents, such as reports, books, or other extensive publications. The function applies in cases where at least 15 % of a document's content is SDG-related. For an SDG to be included in the final label distribution, a given SDG must account for at least 10 % of all SDG-related content within that particular document.

## 2.2. Multilingualism

While a lot of academic publications are either fully in English or have an English abstract, many other texts produced by development practitioners, researchers or universities are not. Texts such as university course descriptions or research project reports tend to be in the local languages. This thus poses additional challenges to any classification attempt.

At the moment, many initiatives offer to identify SDG contents in texts, but all of them work exclusively in English. This causes many problems for many actors, which primarily produce texts in several languages or simply not in English (See section 4 on the applications of the OSDG tool.) To answer this stakeholder need, OSDG team have added support for multiple languages in OSDG 2.0. The system works by using state-of-the-art neural machine translation models to translate the input into English and then pass it through the OSDG workflow.

**Figure 2.** Languages supported by OSDG 2.0

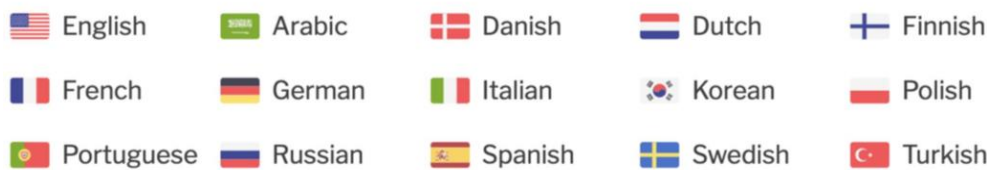

The OSDG team plans adding more languages over time and increase the accessibility of the OSDG ever further.

## 2.3. Front-end of the tool

The illustration below showcases the user-facing interface. In this instance, a user pastes an abstract from a scientific paper and instantly returns the most relevant SDG, in this case SDG 3. If the user does not agree with the result, the 'Suggest SDG labels' functionality allows them to propose a different classification. This submission does not produce any immediate results, but all suggested data are collected to later improve the approach.

**Figure 3.** Example of a successful result for text input.

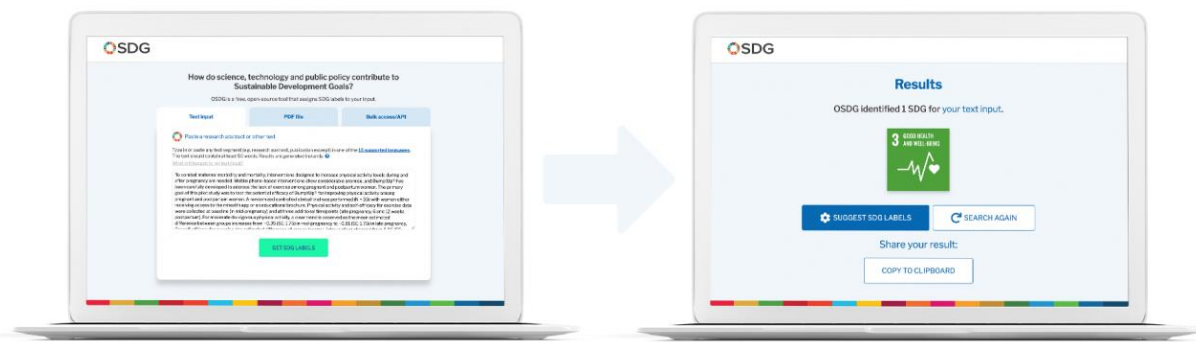

*Source: compiled by the OSDG project team*

Aside from the publicly available tool, the new version can also perform large-scale classification of SDGs through an Application Programming Interface (API). The API works by accepting a user's text input or PDF documents and returning the most relevant SDG labels. The extended tool is freely available for research purposes and can be utilised to assign SDG labels to archives, reports, websites, or more. Examples of working OSDG API applications are presented in section 3.

# 3. Community platform

This section delves deeper into OSDG's approach to training ML models through the process of curating a global citizen science project, the OSDG Community Platform. The data collected via OSDG CP was used to develop sixteen one-vs-rest models for each SDG. While these models are already implemented in OSDG 2.0, they are continuously improved as the exercise remains active.

Since its public launch in March 2021, OSDG CP has aimed to establish the largest and broadest possible public consultation, involving several thousand volunteers. Its principal aim was to recruit participants from highly diverse backgrounds to have adequate coverage of different disciplines and to represent diverse cultural and professional backgrounds, as well as levels of education. The team made it a priority to recruit as many people as possible from the Global South to address the Western bias that existed in OSDG 1.0.

Detailed information about the project format, recruitment process, and project outputs are outlined in the sections below.

## 3.1. Project format

Online volunteers use the platform to participate in labelling exercises through which they validate a text snippet's relevance to an SDG, based on their background knowledge. In each exercise, the volunteer is presented with a short text and an SDG label associated with it, as shown in the figure below. Each exercise session consists of 100 texts of approximately 3 to 6 sentences.

In most cases, the text was provided from sources, such as policies, recent project documents and scholarly publications. Each snippet on the Community platform was validated by up to 9 different volunteers to ensure quality.

**Figure 4.** Example of a labelling session

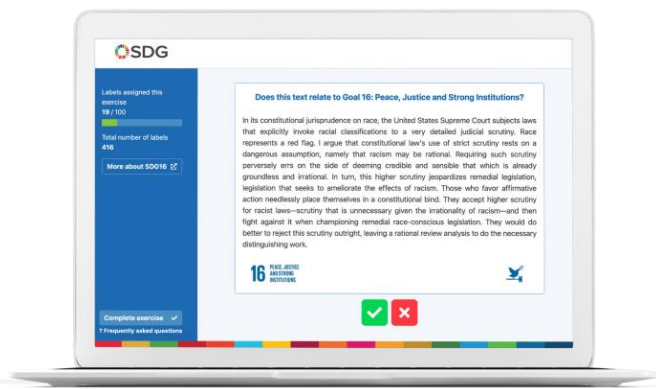

*Source: compiled by OSDG project team*

When developing the approach, the project team aimed to ensure that the process was simple and did not cause unnecessary cognitive load. The simplicity is reflected in the exercise format. For instance, the volunteers only have two choices – to accept or reject the suggested label and are only asked to label one text at a time. If a volunteer needs to verify whether a particular text is related to a particular SDG, all 169 targets along with their indicators are readily available in the sidebar. Finally, although the session consists of 100 texts, the volunteers can stop the exercise at any time without losing any of their progress. The tool also includes stop-points every 20 texts to provide a short break.

The team also provides the ability to customise the exercise, namely by focusing on the preferred SDG area. The first exercise option is to partake in SDG-specific labelling, on a single SDG, e.g. SDG 1 (No Poverty). This SDG-specific exercise is primarily oriented at experts with extensive knowledge of a particular thematic area. Another exercise option exposes participants to a random sequence of texts, where each text could have any SDG as its associated label. Based on overall project statistics, participants are more likely to label a mix of texts, as such sessions appear to be twice as popular than those focusing on a single SDG.

### 3.2. Recruitment

Aside from having direct access to a computer with a working internet connection, the project team provided a few guidelines. The participants were recommended to possess at least a bachelor's degree in any field, advanced English skills, and interest in and/or knowledge of the SDGs.

The project team made sure to communicate the value of the project, highlighting that participants of the project would directly contribute to a better understanding of the SDGs, assist international organisations, academics and companies to advance their research on sustainability, and improve an open-source tool.

The recruitment campaign included advertising on social media and targeted messaging to citizen science platforms. The OSDG founding partner UNDP IICPSD's SDG AI Lab assisted the recruitment efforts by onboarding highly motivated participants through the UN Online Volunteering Platform (OV), currently operating as the Unified Volunteering Platform (UVP). Throughout the active recruitment period of 18 months, the OSDG CP project attracted approximately **900 online UN volunteers**, and nearly **1 000 citizen scientists from other campaigns**. The growth of recruited volunteers is displayed below, along with the rise in assigned SDG labels.

**Figure 5.** OSDG Community Platform participants and labels, per month.

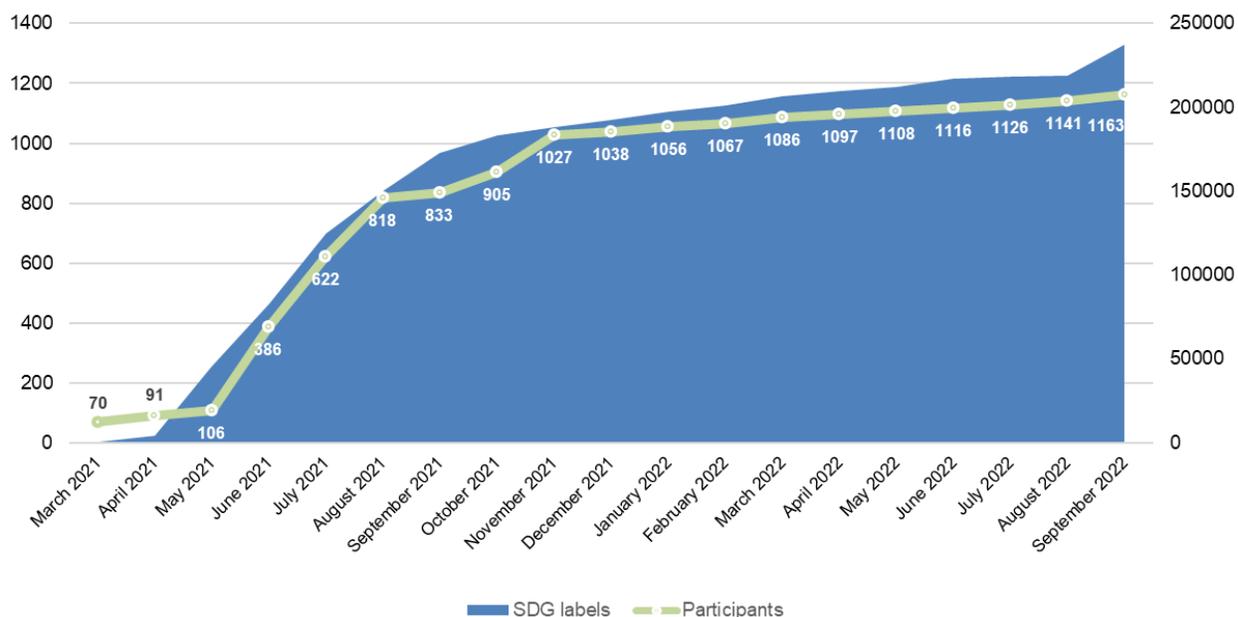

*Source: compiled by the project team*

In providing feedback, several volunteers noted that taking part in the exercise has contributed to their knowledge of specific goals (see figure below).

**Figure 6.** Feedback from the participants

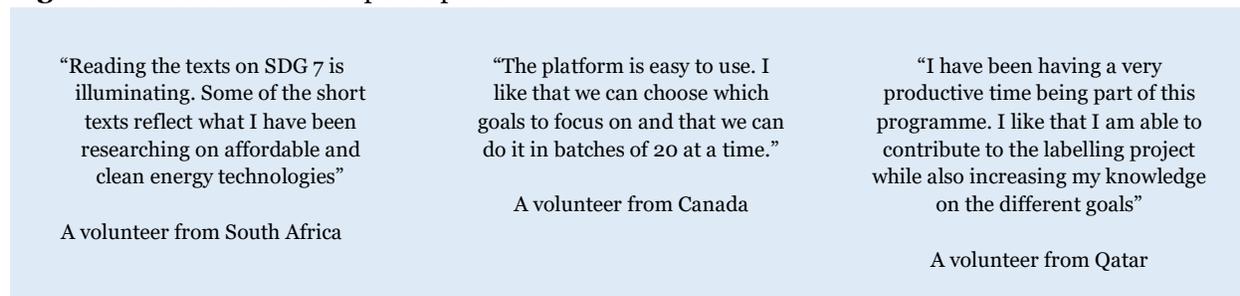

"Reading the texts on SDG 7 is illuminating. Some of the short texts reflect what I have been researching on affordable and clean energy technologies"

A volunteer from South Africa

"The platform is easy to use. I like that we can choose which goals to focus on and that we can do it in batches of 20 at a time."

A volunteer from Canada

"I have been having a very productive time being part of this programme. I like that I am able to contribute to the labelling project while also increasing my knowledge on the different goals"

A volunteer from Qatar

*Source: online exit survey of OSDG CP volunteers*

## 3.3. Quality control

Once registered, each volunteer was invited to take part in a mandatory introductory exercise, which asked them to label 10 pre-selected texts. Afterwards, the volunteer was able to review and compare their answers with other Community members using aggregated statistics. Once the introductory session was complete, the volunteer was able to proceed with the exercise of their choice.

The project team took measures to ensure that newly onboarded citizen scientists would contribute to their maximum capacity without sacrificing the quality of labelled results. The newly onboarded volunteers were invited to weekly webinars, which provided a refresher on the history of SDGs, a step-by-step walkthrough of the exercise, as well as a session to ask questions. The project team also regularly responded to participant queries through a live chat on the platform and email correspondence. Finally, the project team reviewed the labels.

## 3.4. Participant overview

In this section, we briefly overview participants' backgrounds: their countries of residence, age groups, and levels of education.

As of September 2022, the project had achieved an engagement rate of approximately 72 %, meaning that almost 3 out of 4 registered participants have reviewed at least one text snippet. The OSDG Community Platform has at least one active member from 132 countries. 42.9 % of all participants come from Asia, 20.6 % from Africa, and 10.7 % from North America.

**Figure 7.** OSDG Community Platform participants by education level

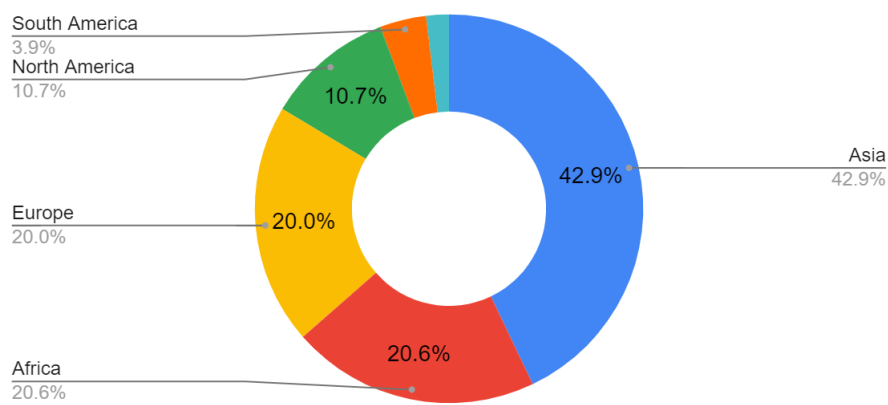

*Source: OSDG CP participant data*

The chart below showcases the diversity of project participants by country. The highest number of participants come from India (14.2 %), followed by the United States (6.2 %), and China (4.2 %).

**Figure 8.** OSDG Community Platform participants by country

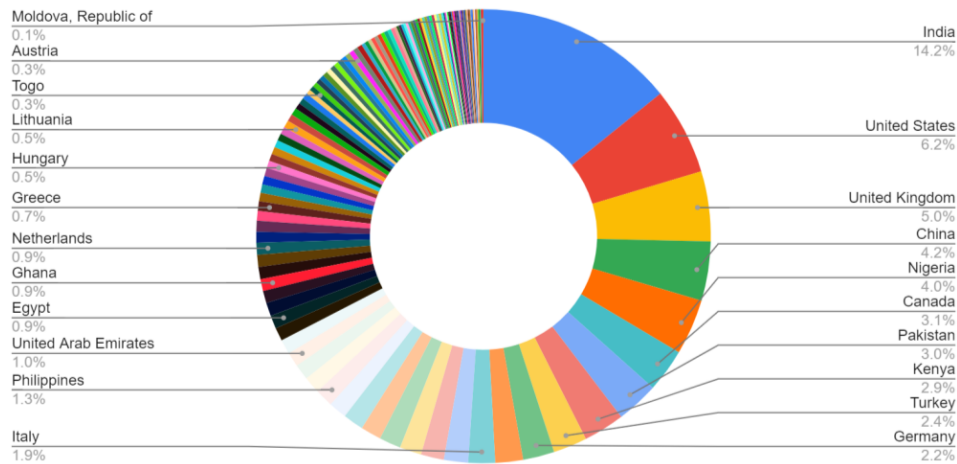

*Source: OSDG CP participant data*

The OSDG CP citizen scientists cover a total of eight different age groups, with the highest number of participants (37.8 %) being between 21 and 25 years, followed by those aged 26-30 (23.8 %), and 31-35 (12.9 %). Less than half of all participants (47.1 %) listed their highest education level to be equivalent to a bachelor's degree, followed by Master's graduates (35.9 %), and high school graduates (10.7 %).

**Figure 9.** OSDG CP participants by age group

**Figure 10.** OSDG CP participants by education level

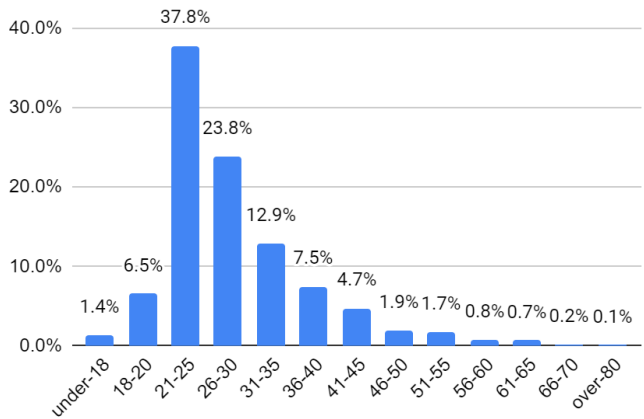
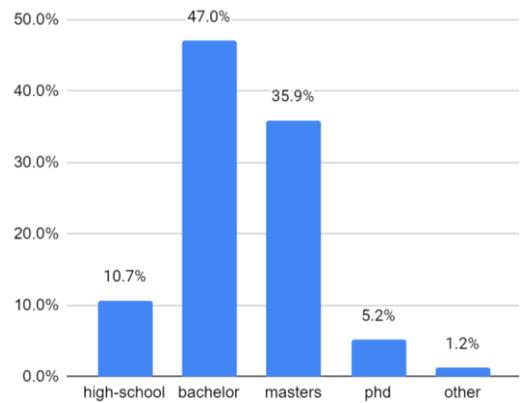

*Source: compiled by the project team*

## 3.5 OSDG Community Dataset

Adhering to citizen science principles, the OSDG team had pledged to give back to the research community and civil society at large in the form of publishing project results. In October 2021, the team released the first edition of the OSDG Community Dataset (OSDG-CD), an SDG-labelled text dataset that can be used to derive insights into the nature of SDGs using either ontology-based or machine learning approaches.

OSDG-CD lists each text snippet and displays the number of volunteers that agreed and disagreed on a specific SDG label, along with the final agreement score. All texts included in the dataset have been validated by at least 3 different volunteers. Currently, the dataset includes texts on 16 SDGs, while the efforts for SDG 17 had been previously halted due to the complexity involved in analysing it. However, the process for including this Goal is underway, and data on it should be published in the forthcoming iterations.

The dataset is publicly available on the Zenodo repository and is updated every quarter (OSDG et al., 2022). As of 1 October 2022, the fifth release contains 37 575 text excerpts, which were validated by the Community volunteers with respect to the 16 SDGs.

These project results have already been well received by researchers and sustainability experts. The crowdsourced data have been applied to make discoveries in research papers, develop machine learning models, and other explorations, presented in more detail in section 3.3.

### 2.4. Future of OSDG-CD

The OSDG Community Platform exercise, along with the collected data, provide valuable feedback on the boundaries of SDG-related research, and the level of consensus among different stakeholders, and contributes to the continuous improvement of the classification tool.

While growth of the dataset is based on volunteer contributions, it will need to be further improved to ensure it responds to new developments. The improvements foreseen include, but are not limited to:
- ensuring full coverage of all SDGs,
- uploading more recent/diverse text content, and
- conceptualising new thematic areas.

## 4. Use cases

Within several years of operation, the OSDG team has identified several use cases for applying its methodology. This section outlines examples of existing applications.

### 4.1. Online classification tool

The open-source classification tool is free to use and does not impose any limits on users uploading texts or PDF files. This method is suitable for users with a relatively small number of texts. It is also frequently used as a verification tool to check if a particular model returns the same SDGs. More recently, the tool was utilised to assess publications by Spanish institutions indexed by the Web of Science. The research team of Sánchez et al. (2022) aimed to inspect whether a particular thematic category addressed topics related to the SDGs and used the online tool to gather these data.

### 4.2. Application Programming Interface (API)

The OSDG team also curates an API, which allows it to automatically assign SDG labels to both text snippets and documents. API access is offered free of charge to scholars conducting research that would be made available to the public in the form of a journal publication, conference paper, etc.
A recent example of the application is the work of Iribarren-Maestro et al. (2022), which aims to compare different strategies to measure the contribution of university research to the SDGs, comparing the perspectives of large publishers with the non-profit proposal of OSDG.

The OSDG API is also functional on several platforms that help to provide additional insights into the SDGs. OSDG has provided API access to an online citizen science hub **SciStarter**, a research affiliate of North Carolina State University and Arizona State University. The platform displays SDG labels as supplementary metadata under each of over 3 000 citizen science projects. This addition allows sustainability-driven citizen scientists to discover projects with relevant global goals and contributes to further mainstreaming of the SDG framework. Another active application of the OSDG API is made available at **SwedishScience**, an open repository of nearly 2 million works published at Swedish universities and authorities. The API tool makes it easy to pinpoint sustainable development topics prevalent in Sweden by reviewing the research abstract of each publication. The API is also deployed at **SDG Careers**, an actively curated SDGs job board. Combined with other labelling methods, the API allows users to filter vacancy announcements by sustainability areas.

**Figure 11.** OSDG API implementation at *SciStarter* and *SwedishScience*

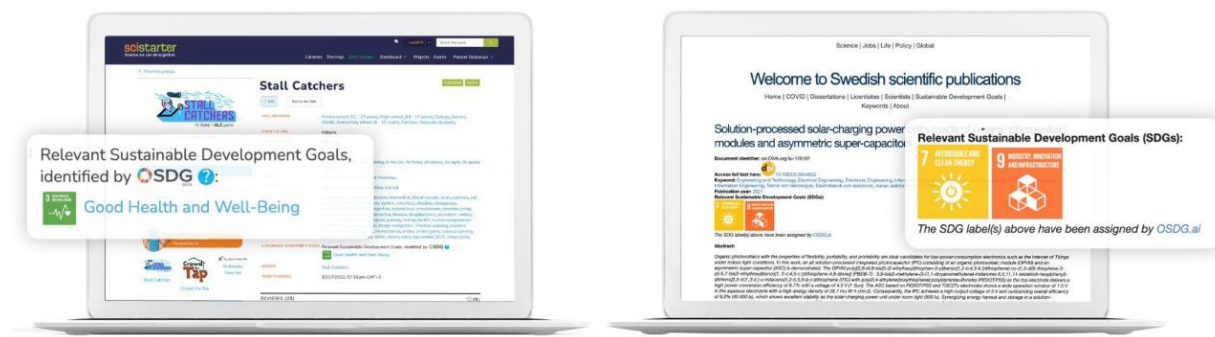

*Source: compiled by the project team, images from the Stall Catchers project and SwedishScience.se*

## 4.3. OSDG Community Dataset

Another use case relates to the OSDG Community Dataset, a resource of SDG data, updated every quarter. The dataset has already been used to derive insights into the nature of SDGs by entities such as the University of Hong Kong. The university's Technology-Enriched Learning Initiative (TELI) and Common Core Office aimed to evaluate SDG-related formal learning activities in the university's common core curriculum. The dataset was used to train the model and test it for more than 100 courses (Lei et al., 2022). OSDG-CD was also utilised in a study that aimed to link energy-related policies of the European Green Deal to the SDGs using ML techniques (Koundouri et al., 2022). The authors extended the manual linkage of policy texts to the SDGs by using ML and Natural Language Processing (NLP) to automate it and ended up using the OSDG dataset to fine-tune a pre-trained BERT model. The OSDG team receives regular inquiries from sustainability enthusiasts, data scientists, students and NGOs interested in using the dataset, further demonstrating the growing importance of SDG research and the demand for such data.

## 4.4. Bespoke research

The OSDG team regularly performs large-scale SDG assessments for research or academic entities. The growing number of requests shows the increased awareness and understanding of SDG impact and a need to derive insights from existing data. The team's growing portfolio is a direct reflection of the demand to assess the extent to which SDGs are promoted within university curricula, publications, research projects and even university events.

The project team has provided SDG analysis for the University College London (UCL), further detailed in the university's SDGs Report 2020-21 (UCL, 2021). The OSDG methodology was employed to classify the descriptions of more than 6 000 modules in the university's online catalogue. The OSDG can also benefit universities whose data, such as course descriptions or research repositories, are in multiple languages. Thanks to the tool's translation capabilities, the project has already implemented pilot case studies with the Católica Lisbon School of Business & Economics, part of the Catholic University of Portugal. A similar exercise is being carried out with York University (Canada). The API is being applied to help identify SDG labels for York University courses based on the course descriptions in French or English.

**Figure 12.** SDGs-related teaching at UCL, as assessed by OSDG

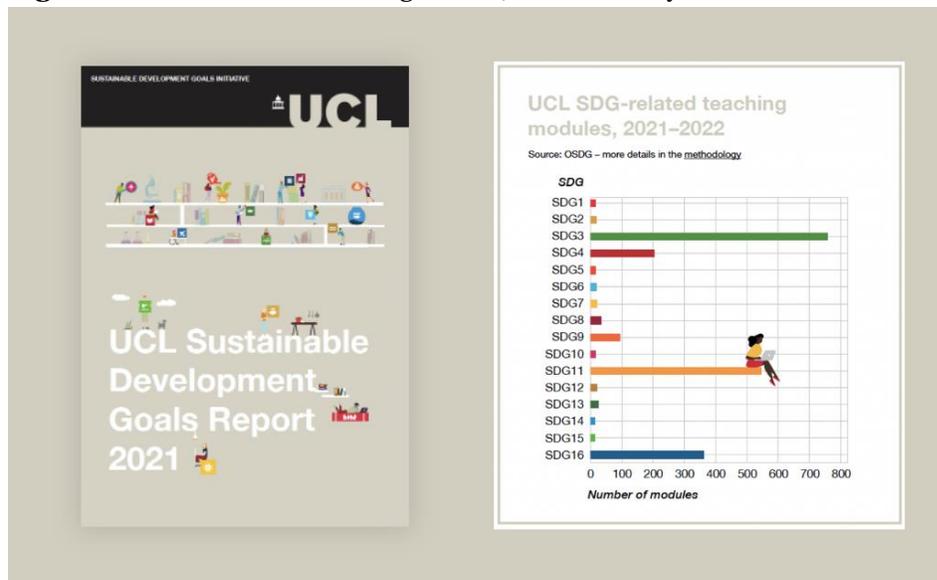

*Source: [How we measured UCL's SDG-related activity](#), UCL.*

# Conclusions

## Request for feedback and contributions

The OSDG team welcomes all contributions to further improve the tool. Researchers are invited to suggest new data sources, improve the current procedures for data cleaning or processing, or add new items to the existing ontology. This can be done by creating a pull request on GitHub. The team also welcomes any feedback regarding the perceived quality of the labels returned for text snippets or PDF files, including commentary on the translation service. Finally, the team is actively looking into expanding coverage of thematic areas that are not sufficiently recognised by the SDG framework. This feedback adds to the discussion on what exactly is a contribution to the SDGs.

As part of the effort to apply the OSDG know-how, the team remains open to further collaboration in developing the tool. For all queries and feedback, please contact info@osdg.ai.

# Links

OSDG tool: https://www.osdg.ai/

OSDG Community Platform: https://www.osdg.ai/community

OSDG on GitHub: https://github.com/osdg-ai/